# Ferromagnetic Cr$_4$PtGa$_{17}$: A Half-Heusler-Type Compound with a Breathing Pyrochlore Lattice

*Xin Gui,[a] Erxi Feng,[b] Huibo Cao,[b] and Robert J. Cava [a]\**

[a] Department of Chemistry, Princeton University, Princeton, NJ 08540, USA
[b] Neutron Scattering Division, Oak Ridge National Laboratory, Oak Ridge, TN 37831, USA

## *Abstract*

We describe the crystal structure and elementary magnetic properties of a previously unreported ternary intermetallic compound, Cr$_4$PtGa$_{17}$, which crystallizes in a rhombohedral unit cell in the noncentrosymmetric space group *R*3*m*. The crystal structure is closely related to those of XYZ half-Heusler compounds, where X, Y and Z are reported to be single elements only, occupying three different face-centered cubic sublattices. The new material, Cr$_4$PtGa$_{17}$, can be most straightforwardly illustrated by writing the formula as (PtGa$_2$)(Cr$_4$Ga$_{14}$)Ga (X=PtGa$_2$, Y = Cr$_4$Ga$_{14}$, Z = Ga), that is, the X and Y sites are occupied by clusters instead of single elements. The magnetic Cr occupies a breathing pyrochlore lattice. Ferromagnetic ordering is found below T$_C$ ~61 K, by both neutron diffraction and magnetometer studies, with a small, saturated moment of ~0.25 μ$_B$/Cr observed at 2 K, making Cr$_4$PtGa$_{17}$ the first ferromagnetically ordered material with a breathing pyrochlore lattice. A magnetoresistance of ~140% was observed at 2 K. DFT calculations suggest that the material has a nearly-half-metallic electronic structure. The new material, Cr$_4$PtGa$_{17}$, the first realization of both a half-Heusler-type structure and a breathing pyrochlore lattice, might pave a new way to achieve novel types of half-Heusler compounds.

*Introduction*

Heusler compounds are a large family of materials, comprising over 1500 members. Extensive studies have been carried out on their crystal structures and intriguing properties since the discovery of Cu$_2$MnAl more than a century ago.[1–3,9] In terms of crystal structure, they are generally categorized into two subfamilies: the Heusler structure and the half-Heusler (HH) structure. The former has general formula X$_2$YZ and crystallizes in the centrosymmetric cubic space group $Fm\bar{3}m$ (No. 225) where X, Y and Z atoms occupy the 8$c$ (¼, ¼, ¼), 4$a$ (0, 0, 0) and 4$b$ (½, ½, ½) sites respectively. The latter has general formula XYZ, adopting the noncentrosymmetric cubic space group $F\bar{4}3m$ (No. 216). The X, Y and Z atoms in HH compounds occupy the 4$a$ (0, 0, 0), 4$b$ (½, ½, ½) and 4$c$ (¼, ¼, ¼) sites respectively. The Heusler structure consists of four interpenetrating face-centered cubic (*fcc*) sublattices while there are three interpenetrating *fcc* sublattices in the HH structure.[2,3] To date, X, Y and Z in both structure types are reported to be single elements, where Z has the largest electronegativity and X is the most electropositive.[3] Many attractive magnetism-related properties have been found and investigated for Heusler and half-Heusler compounds, including, for example, ferromagnetic (FM) and anti-ferromagnetic (AFM) ordering.[4–9] Half-metallicity in ferromagnetic/ferrimagnetic materials,[10–12] which can be applied in spintronics due to its resulting high degree of spin polarization, was predicted and confirmed in half-Heusler compounds such as NiMnSb, while many Heusler compounds behave similarly.[13–19] Heusler and half-Heusler compounds are also known to be superconductors,[20–26] thermoelectric materials,[27–33] shape-memory materials,[34–38] and topological insulators.[26,39–42]

Research on geometrical frustration in spinels and pyrochlores, which have the general AB$_2$X$_4$ and A$_2$B$_2$X$_7$ formulas, respectively, has also been of significant recent interest.[43–52] The B atoms in these structures occupy the centers of BX$_6$ octahedra, and by themselves construct a corner-sharing framework of tetrahedra known as the pyrochlore lattice.[43,52] The pyrochlore lattice can be distorted, in one case becoming a three-dimensional framework with alternating larger and smaller tetrahedra instead of identical ones, known as a "breathing pyrochlore lattice". Such systems are currently being explored, both theoretically and experimentally.[52–58] Two examples of the breathing pyrochlore lattice materials are the spinel-based compounds LiGaCr$_4$O$_8$ and LiInCr$_4$O$_8$.[53] Both compounds antiferromagnetically order near 15 K.[53] As indicated by its name, the breathing pyrochlore lattice can also occur in pyrochlore-related compounds such as $Ba_3Yb_2Zn_5O_{11}$.[55,58] No evidence of magnetic ordering has been found above ~0.38 K in this material.

Since Heusler/half-Heusler compounds and compounds with a breathing pyrochlore lattice have significant impact, expanding this family of materials is of interest. Here we report the synthesis of single crystals of a previously unreported ternary intermetallic compound, $Cr_4PtGa_{17}$, which crystallizes in the noncentrosymmetric trigonal space group $R3m$ (No. 160). Even though it does not adopt a cubic unit cell, the crystal structure can be interpreted as a complex half-Heusler type, with the formula written as $(PtGa_2)(Cr_4Ga_{14})Ga$. The Cr atoms in the structure construct a framework where four Cr atoms form a tetrahedron and all tetrahedra are corner-shared with other tetrahedra, i.e., they form a breathing pyrochlore lattice. We have investigated the magnetic properties and electronic structure of this material and it appears to be the first reported ferromagnetically ordered material with a breathing pyrochlore lattice. The new material, with the first realization of combination of half-Heusler structure and breathing pyrochlore magnetic framework, provides a unique platform for investigating the magnetic properties of materials.

## *Experimental Details*

**Single Crystal Growth:** The self-flux method was employed to obtain single crystals of $Cr_4PtGa_{17}$. Cr powder (99.95%, ~200 mesh, Alfa Aesar), Pt powder (>99.98%, ~60 mesh, Alfa Aesar) and Ga (99.999%, ingot, Alfa Aesar) were placed in an alumina crucible in the molar ratio of 4:1:40. The crucible was then placed into an evacuated silica tube. Quartz wool and some glass pieces were put above the crucible to facilitate the separation of the Ga flux from the crystals after cooling. The tube was sealed under vacuum (<5×10$^{-3}$ Torr) after purging by argon. It was heated in a box furnace to 700 °C and held for 24 hours followed directly by another heat treatment at 1050 °C for two days. To ensure good crystallization, the furnace was slowly cooled to 500 °C at the rate of 3 °C per hour. The Ga flux was centrifuged out at this temperature. Crystals with dimensions up to 4×4×1 mm$^3$ with triangular faces were obtained.

**Structure and Phase Determination:** More than ten $Cr_4PtGa_{17}$ crystals (~50×50×10 μm$^3$) were studied by single crystal X-ray diffraction to determine the crystal structure of the new material. The structure, consistent among all crystals, was determined using a Bruker D8 VENTURE diffractometer equipped with APEX III software and Mo radiation ($\lambda_{K\alpha}$= 0.71073 Å) at room temperature. The crystals were mounted on a Kapton loop protected by glycerol. Data acquisition was made *via* the Bruker SMART software with corrections for Lorentz and polarization effects included. A numerical absorption correction based on crystal-face-indexing was applied through the use of *XPREP*. The direct method and full-matrix least-squares on $F^2$ procedure within the SHELXTL package were employed to solve the crystal structure.[59,60] Powder Xray diffraction patterns, obtained with a Bruker D8 Advance Eco with Cu Kα radiation and a LynxEye-XE detector, were employed to determine the phase purity of the crystals whose properties were studied. The patterns were fitted by the Rietveld method in Fullprof using the crystal structure obtained from the single crystal data.

**Physical Property Measurement:** The DC magnetization was measured from 2 to 300 K under various applied magnetic fields using a Quantum Design Dynacool Physical Property Measurement System (PPMS), equipped with a vibrating sample magnetometer (VSM) option. The magnetic susceptibility was defined as M/H. Field-dependent magnetization data was collected at different temperatures with applied magnetic fields ranging from -9 T to 9 T. The resistivity measurements were carried out in in the same system using the four-probe method between 1.8 K to 300 K under magnetic fields up to $\mu_0 H = 9$ T. Platinum wires were attached to the samples by silver epoxy to ensure ohmic contact.

**Single Crystal Neutron Diffraction:** Single crystal neutron diffraction characterization was performed on the DEMAND diffractometer (HB-3A) at the High Flux Isotope Reactor (HFIR) at Oak Ridge National Laboratory (ORNL).[61] A crystal of ~4×4×1 mm$^3$ was selected and mounted on an aluminum pin. Measurements were carried out between 5 K and 80 K in a closed-cycle refrigerator with an incident neutron wavelength of 1.542 Å from a bent Si-220 monochromator. The nuclear and magnetic structure refinements were carried out using the FullProf refinement Suite.[62]

**Electronic Structure Calculations:** The electronic structure and electronic density of states (DOS) of $Cr_4PtGa_{17}$ were calculated using the WIEN2k program, which employs the full-potential linearized augmented plane wave method (FP-LAPW) with local orbitals implemented.[63,64] The electron exchange-correlation potential used to treat the electron correlation was the generalized gradient approximation.[65] The conjugate gradient algorithm was applied, and the energy cutoff was set at 500 eV. Reciprocal space integrations were completed over a 6×6×6 Monkhorst-Pack *k*-point mesh.[66] Spin-orbit coupling (SOC) effects were only applied for the Pt atom. Spin-polarization (ferromagnetism with the moment oriented in the (010) direction) was only employed for the Cr atom. The structural lattice parameters obtained from experiment were used in all calculations. With these settings, the calculated total energy converged to less than 0.1 meV per atom.

## *Results and Discussion*

**Crystal Structure and Phase Determination:** $Cr_4PtGa_{17}$ crystallizes in the trigonal space group *R3m* (No. 160). The crystallographic data, including atomic positions, site occupancies, and refined anisotropic displacement parameters (and equivalent isotropic thermal displacement parameters) are listed in Tables 1, 2 and 3. The structure is non-centrosymmetric – the absolute structure parameter (Flack parameter) was determined to be 0.01(1), indicating that the absolute structure determined, shown here in Figure 1(a), is correct. During the process of structure refinement, crystal structure solutions were attempted in other centrosymmetric and noncentrosymmetric space groups, however, none of them gave reasonable refinement results. As shown in the main panel of Figure 1(f), the Rietveld fitting parameters for crystals used for the magnetic characterization, crushed into powders, are $R_p$ = 3.77%, $R_{wp}$ = 5.26%, $R_{exp}$ = 3.51% and $\chi^2$ = 2.25. The low values reveal a high-quality fit and a high phase purity. Finally, the crystal structure determined by single crystal X-ray diffraction was supported by the single crystal neutron diffraction experiments. As can be seen in Figure 1(a), $PtGa_6$ octahedra occupy the vertices of the trigonal unit cell. The Cr atoms, on the other hand, are coordinated by nine Ga atoms, forming $CrGa_9$ triply-capped trigonal prisms, as shown in Figure 1(e). In this polyhedron, three Ga atoms can be regarded as constructing a $Ga_3$ triangle with three of these $Ga_3$ triangles stacked in a staggered way and bonded with Cr atoms. $Cr_4PtGa_{17}$ crystals are typically triangular plates, with a representative one shown in the inset of Figure 1(f).

**The Breathing Pyrochlore Lattice in $Cr_4PtGa_{17}$:** A view of the crystal structure of a representative $A_2B_2O_7$ oxide pyrochlore, (space group $Fd\bar{3}m$), is shown in Figure 2(a). The A-site pyrochlore lattice extracted from this structure (both the A and B sublattices are pyrochlore-type lattices), is presented in Figure 2(b) where the corner-shared tetrahedral framework of A atoms can be found. The pyrochlore lattice can be viewed as two layers of Kagome planes bridged by a layer of atoms positioned appropriately in the middle. Therefore, all tetrahedra in the pyrochlore lattice are identical, i.e., $d_1 = d_2 = d_\Delta$, where $d_1$ and $d_2$ are the separations of the A atoms between Kagome planes and the bridging atom, and $d_\Delta$ is the A-A bond length within the Kagome plane. A similar lattice can be found in $Cr_4PtGa_{17}$, as shown in Figure 2(c). Two Kagome planes are illustrated in the figure: the triangular lattice of bridging Cr atoms is located closer to one of the Kagome planes than the other. Thus, the larger and smaller tetrahedra stack in alternate planes to form a breathing pyrochlore lattice: $d_1$ (4.999 (2) Å) and $d_{\Delta 1}$ (5.002 (2) Å) are longer than $d_2$ (3.026 (2) Å) and $d_{\Delta 2}$ (3.024 (1) Å). $d_1$ and $d_2$ are very similar to $d_{\Delta 1}$ and $d_{\Delta 2}$ but are not identical due to the rhombohedral symmetry. The possible structural, thermodynamic and crystal chemical aspects of breathing pyrochlore lattice formation have been described earlier[67].

**Relationship between the Crystal Structure of $Cr_4PtGa_{17}$ and Half-Heusler Compounds:** Half-Heusler (HH) compounds, typically consisting of three different elements, XYZ, crystallize in a cubic unit cell with space group $F\bar{4}3m$ (No. 216) where each element constructs a face-centered cubic (*fcc*) sublattice and interpenetrates the other two *fcc* sublattices (see the right figure in Figure 1(b)). In a single unit cell of HH compounds, Z atoms are found at the cell corners and face centers, the Y atoms occupy the centers of the cell edges while the X atoms occupy half of the centers of simple cubes constructed by four Y and four Z atoms in a tetrahedral pattern. In $Cr_4PtGa_{17}$, although the atomic ratio does not obey the 1:1:1 ratio typically seen for HH compounds, a pseudo-cubic cell can still be found, as shown in Figure 1(b) (Left). The pseudo-cubic cell of $Cr_4PtGa_{17}$ is distorted from the regular cubic cell by an angle of only ~0.009(1)°, i.e., the angle between the three pseudo-cubic axes of the pseudo cubic cell is 90° ± 0.009(1)° rather than exactly being 90 degrees. (The Ga atoms bonded to the Cr and Pt atoms are omitted from the representation in Figure 1b for clarity.). The corners and face centers of the pseudo-cubic cell (i.e. the *fcc* positions) are occupied by Ga atoms - the Z site in HH compounds. The geometric center of the larger of the $Cr_4$ tetrahedra of the breathing pyrochlore lattice occupies the center of the edge of the pseudo-cubic cell, i.e., the Y site in HH compounds, while four $PtGa_6$ octahedra are found in a site similar to that occupied by a single X atom in HH compounds. Figures 1(c) and (d) illustrate the *fcc* sublattices formed by the $PtGa_6$ octahedra and $Cr_4$ tetrahedra. Therefore, the crystal structure of $Cr_4PtGa_{17}$ is strongly related to the HH structure, with its formula interpreted as $(PtGa_2)(Cr_4Ga_{14})Ga$, XYZ, X= $PtGa_2$, Y= $Cr_4Ga_{14}$, Z= Ga, i.e., X and Y, which are single elements in HH compounds, are occupied by clusters in $Cr_4PtGa_{17}$.

**Magnetic Characterization:** Magnetic property measurements were performed on triangular $Cr_4PtGa_{17}$ single crystals. The plate of the crystal was identified by single crystal X-ray diffraction as being the plane perpendicular to the (001) direction (the *c* axis). All magnetic data were collected for both directions, parallel to the *c* axis (H//*c*) and perpendicular to the *c* axis (H⊥*c*), between 300 and 2 K. Both zero-field cooling (ZFC) and field-cooling (FC) modes were applied.

The results of the magnetic characterization of the crystals are summarized in Figure 3. According to Figure 3(a) and (b), under an applied magnetic field of 0.3 T, Curie-Weiss behavior is observed for both directions by fitting with a modified Curie-Weiss (CW) law with $\chi_0$: $\chi = \chi_0 + C/(T-\theta_{CW})$ where $\chi$ is magnetic susceptibility of the material, $\chi_0$ and C are independent of temperature (the latter related to the effective moment and the former to the core diamagnetism and temperature independent paramagnetic contributions such as Pauli paramagnetism), and $\theta_{CW}$ is the Curie-Weiss constant. When H//*c*, $\theta_{CW}$ is 69.3 (6) K while it is 66.9 (7) K when H⊥*c*. Positive Curie-Weiss constants indicate the the dominance of ferromagnetic interactions in $Cr_4PtGa_{17}$. The effective moment was obtained from the CW fitting by $\mu_{eff} = \sqrt{8C}$. When H//*c*, it is determined to be ~1.91 $\mu_B$/f.u., i.e., ~0.48 $\mu_B$/Cr while when H⊥*c*, $\mu_{eff}$ is ~2.20 $\mu_B$/f.u., i.e., ~0.55 $\mu_B$/Cr. All the $\mu_{eff}$ obtained from the CW fits are smaller than what is expected for isolated magnetic atoms (for the minimum case, one localized electron per atom, spin ½ per magnetic atom, 1.73 $\mu_B$/magnetic atom, is expected). Based on the crystal structure, however, even in a localized spin picture, every four Cr atoms can be considered as being part of a $Cr_4$ tetrahedron due to the short Cr-Cr distances ($d_2$ = 3.026 (2) Å and $d_{\Delta 2}$ = 3.024 (1) Å) in the small $Cr_4$ tetrahedron in the breathing pyrochlore lattice. It is plausible to speculate that in a localized picture each $Cr_4$ tetrahedron can be treated as a metal cluster that displays an effective electronic state that leads to between 1 and 2 unpaired electrons per cluster.[68] By looking at the low-temperature (<100 K) data, ferromagnetic behavior can be seen when H⊥*c*, that is, that $\chi$ increases below the Curie temperature ($T_C$) and reaches a plateau at lower temperature. However, a peak can be seen at ~43 K when H//*c* which might originate from the movement of domain walls. To better probe the magnetic behavior below $T_C$, $\chi$ *vs* T curves under low magnetic fields were collected. As shown in Figure 3(c) and (d), clear ferromagnetic behavior can be observed for both directions. The insets present the first derivatives of $\chi$ *vs* T under different magnetic fields, which demonstrate that a similar temperature (~60 K) for where the material shows non-CW (non-linear) behavior is seen for both directions. Moreover, the fact that this temperature in the $d\chi/dT$ curves is similar to the Curie-Weiss constant reveals that there might not be strong magnetic frustration in the material. With stronger magnetic field, the transition becomes broader which might be due to the fact that the saturated ferromagnetically ordered spins at higher field, i.e., 200 Oe, need higher energy, i.e., more heating, to enter the paramagnetic state than the unsaturated ferromagnetically ordered spin at lower field, i.e., 20 Oe. When H//$c$, for magnetic fields above 100 Oe, kinks can be found just below 60 K which can be attributed to the movement of domain walls.

Hysteresis loops from -9 T to 9 T are shown in Figure 3(e) and (f), which illustrate the typical behavior expected for a ferromagnetic material. Ferromagnetism is frequently observed in Cr intermetallic compounds, such as for the two-dimensional (2-D) ferromagnets $Cr_2Ge_2Te_6$[68-70] and $CrI_3$[71-73]. The magnetization saturates after a small magnetic field, ~145 Oe for H⊥$c$ and ~160 Oe for H//$c$. Small coercive fields for both directions (~20 Oe for H⊥$c$ and ~25 Oe for H//$c$), as can be seen in the insets, reveal the soft ferromagnetic character of $Cr_4PtGa_{17}$. The saturated moment for H⊥$c$ is ~0.94 $\mu_B$/f.u., i.e., ~0.24 $\mu_B$/Cr, while it is ~1.00 $\mu_B$/f.u., i.e., ~0.25 $\mu_B$/Cr, when H//$c$. Therefore, either isotropic behavior or weak magnetic anisotropy can be found in $Cr_4PtGa_{17}$. Small saturated moments have also been observed in other materials such as $CrBe_{12}$.[74] Moreover, magnetization at 2 K, when H//$c$, increases at magnetic fields of ~ 5 T.

**Single Crystal Neutron Diffraction:** To better interpret the magnetic properties of $Cr_4PtGa_{17}$, single crystal neutron diffraction was carried out at Oak Ridge National Laboratory. The nuclear fit to the neutron diffraction data shows good consistency with single crystal X-ray diffraction, supporting the conclusion that $Cr_4PtGa_{17}$ crystallizes in rhombohedral unit cell with space group *R*3*m*. The magnetic refinement of the neutron diffraction data leads to in-plane ferromagnetic ordering of the Cr atom moments, as shown in Figure 4(b). By comparing the nuclear peak intensity and magnetic peak intensity for Bragg peak (0 0 3) of two different crystals, as shown in Figure 4(a), the magnetic moment was determined to be ~0.23 $\mu_B$/Cr, that is, ~0.92 $\mu_B$/f.u., consistent with what is observed in field-dependent magnetization measurements.

As shown in Figure 4(c), the ordering parameter measured between 4 K and 80 K at a propagation vector of (0 0 3) clearly shows that magnetic ordering becomes detectably strong beginning at around 60 K. This temperature is determined by the intersection of a linear fit (60-80 K) and a power-law fit (4-60 K) to the intensity of the peak, which can have both structural and magnetic contributions for a ferromagnet, using the function: $I = A(\frac{T_M-T}{T_M})^{2\beta} + B$, where $T_M$ is the magnetic phase transition temperature, A is a constant, B is the background and β is the critical exponent of the order parameter.[75] $T_M$ from the fit is 61(2) K, consistent with the transition temperature determined from other data, and β is 0.3 (2) consistent with what is shown in Figure 3.

**Resistivity and Magnetoresistance:** The temperature-dependence of the electrical resistivity is important to determine how the charge carriers are sensitive to the magnetic state of the material and also to characterize the crystal quality through the residual-resistivity-ratio (RRR). In addition, materials with large magnetoresistance, which is based on the magnetic-field-dependence of the resistivity at various temperatures, have the potential to be applied in multiple areas, such as magnetic random-access memory (MRAM) and hard disk drives.[76,77] To better understand the electrical properties, then, fundamental resistivity and magnetoresistance measurements were carried out on $Cr_4PtGa_{17}$ single crystals. Figure 5(a) presents the temperature-dependence of the resistivity under no magnetic field. The resistivity decreases monotonically with decreasing temperature, implying metallic behavior. The RRR ($\rho_{300 K}/\rho_{2 K}$) is ~233, which indicates high crystal quality and a low degree of defect scattering. At ~60 K, a kink can be observed in both curves - consistent with the magnetic ordering temperature obtained in the magnetic data. The transition is clearly observed by looking at the first derivative of the ρ vs T curves, as shown in the inset of Figure 5(a). The low-temperature resistivity (2 to 50 K) was then fitted by the following equation for a ferromagnetic metallic system:

$$\rho(T) = \rho_0 + \rho_M(T) + \rho_P(T)$$

where $\rho_0$ is the residual resistivity, $\rho_M(T) = AT^2$ where A is the magnon scattering strength and

$$\rho_P(T) = B\left(\frac{T}{\theta_D}\right)^5 \int_0^{\frac{\theta_D}{T}} \frac{x^5}{(e^x-1)(1-e^x)} dx$$

where B is the phonon scattering exponent and $\theta_D$ is the Debye temperature. The fitted $\rho_0$ for the zero-field cooled curve is 8.3 (1) μΩ cm and $\theta_D$ is 43(1) K. Moreover, since half-Heusler compounds can be half-metallic, it is potentially informative to include a half-metallic term in $\rho_M(T)$, thus, $\rho_M(T) = AT^2 e^{(-\Delta/T)}$ where Δ is the energy gap between the Fermi level and the band edge of the minority spin carriers. In Figure 5(a), Δ was fitted to be -5.0 (3) K, indicating that the Fermi level is below the band edge of the minority spin carriers; no bandgap can be seen according to the resistivity curve fitting.

Figure 5(b) presents the magnetoresistance from 0 to 9 T for $Cr_4PtGa_{17}$ crystals with the magnetic field applied perpendicular to the (001) plane. At 2 K, $Cr_4PtGa_{17}$ shows a MR% of about 140%, which decreases with increasing temperature. The MR% increases with a larger slope at 2 K and 10 K below

~1.5 T, which can be attributed to the alignment of the spins in different magnetic domains in a ferromagnet. Below ~1.5 T, as described in Figure 3(e) & 3(f), at 2 K, the spins are not totally aligned by the field, i.e., they are not saturated. When the magnetic field increases, the spins become totally aligned and thus, the magnetoresistance increases with a smaller slope due to the decrease of scattering from different magnetic domains. The 140% magnetoresistance at 2 K for $Cr_4PtGa_{17}$ is about half of that for another half-Heusler compound, ScPtBi,[78] but is much smaller than what has been reported for materials such as $WTe_2$.[79] The MR is negative at 50 K, decreasing until $\mu_0H$ reaches ~ 4 T, which might be related to the change of carrier density.[80,81]

**Electronic Structure:** The electronic structure of $Cr_4PtGa_{17}$ was calculated as described. The results include both the band structure and the electronic density of states (DOS). Figure 6(a) & 6(b) present the band structure and DOS from -2 eV to 2 eV with and without consideration of spin-orbit coupling (SOC) effects on the Pt atoms while Figure 6(d) illustrate the Brillouin Zone (BZ) and wavevector path used in the calculations. The contributions from the dominant orbitals of the different types of atoms are also shown in the DOS plots. The calculations reveal a metallic behavior for this material, with a bandgap found above Fermi level - $E_F$ is located just below the top edge of the valence band. For the case without SOC, a DOS peak can be seen at $E_F$, the majority of which can be attributed to the 3$d$ states of the Cr atoms. Similar behavior can be found when including SOC - 3$d$ states of Cr are found to be dominant in the exhibited energy range. Moreover, band splitting can clearly be observed with inclusion of SOC and the van Hove singularity around $E_F$ becomes sharper. A gap is opened up just below $E_F$ at the $\Gamma$ point after inclusion of SOC, as shown by the red arrows in Figure 6(a) & 6(b), leading to a continuous small bandgap along a closed wavevector route $\Gamma$-L-P-Z-$\Gamma$ a few meV below $E_F$. The energy gaps above $E_F$ obtained in the band structure calculations and the DOS calculations are 0.65 eV and 0.57 eV with SOC included. This minor difference is within the error of the methods.

After introducing spin polarization on the Cr atoms due to the ferromagnetism, as shown in Figure 6(c), there appears to be metallic behavior in one spin orientation. There is a semiconducting-like bandgap for the other spin orientation; the corresponding DOSs for both spin orientations present no bandgap for this material. For the spin-up direction, a very low density of states (~0.12 states/eV) at $E_F$ can be found, as shown in the inset of Figure 6(c). For comparison, the calculated DOS at $E_F$ for the spin-down orientation is ~25.15 states/eV, which can be attributed to $d$ states of Cr, and it is over 200 times that of that in spin-up case. The calculated magnetic moment is 0.96 $\mu_B$/f.u., which is consistent with what is observed in the magnetic measurements (~0.94 $\mu_B$/f.u. for H⊥$c$ and ~ 1.00 $\mu_B$/f.u. for H//$c$). The results of the calculation indicate that the new material possesses a near-half-metallic electronic structure where the spin is not fully polarized at the Fermi level, which is consistent with the negative energy gap fitted from the half-metallic resistivity term as described in the resistivity section above. Moreover, considering that the magnetic moment of the theoretical calculation is set to be oriented in the (010) direction, the consistent results of near-half-metallic behavior and saturated moment between the calculations and experiments further confirm the in-plane magnetic structure obtained by neutron single crystal diffraction.

## *Conclusions*

We report a new ternary intermetallic compound, $Cr_4PtGa_{17}$, crystallizing in a rhombohedral unit cell in the noncentrosymmetric space group $R3m$. The formula can be written as $(PtGa_2)(Cr_4Ga_{14})Ga$ to illustrate the relationship of this structure with those of half-Heusler compounds. A breathing pyrochlore lattice can also be found in the new material. Thus, it makes $Cr_4PtGa_{17}$ the first reported material with both the half-Heusler structure type and a breathing pyrochlore lattice. Based on the physical property measurements and neutron diffraction, the new material was determined to order ferromagnetically below $T_C$ ~ 61 K with a small saturated moment ~0.25 $\mu_B$/Cr, making this the first ferromagnetically ordered material with a breathing pyrochlore lattice. A magnetoresistance of ~140% can be observed at 2 K. DFT calculations indicate that $Cr_4PtGa_{17}$ possesses a near-half-metallic electronic structure. This half-Heusler-like material with a breathing pyrochlore lattice may provide a novel platform for investigating ferromagnetic materials, especially because single crystals of substantial volume can be grown.

## *Author Information*

Corresponding Author: rcava@princeton.edu

Notes: The authors declare no competing financial interest.

## *Acknowledgements*

X. G. and R. J. C. were supported by the Gordon and Betty Moore Foundation, EPIQS initiative, grant GBMF-9006. E.F. and H.C. acknowledge the support of U.S. DOE BES Early Career Award No.

KC0402020 under Contract No. DE-AC05-00OR22725. This research used resources at the High Flux Isotope Reactor, a DOE Office of Science User Facility operated by the Oak Ridge National Laboratory.## References

(1) Heusler, F.; Starck, W.; Haupt, E. Verh. Magnetisch-chemische studien. *DPG* **1903**, *5*, 220.

(2) Wollmann, L.; Nayak, A. K.; Parkin, S. S. P.; Felser, C. Heusler 4.0: Tunable Materials. *Annu. Rev. Mater. Res.* **2017**, *47*, 247–270.

(3) Graf, T.; Felser, C.; Parkin, S. S. P. Simple Rules for the Understanding of Heusler Compounds. *Prog. Solid. State Ch.* **2011**, *39*, 1–50.

(4) Mohn, P.; Blaha, P.; Schwarz, K. Magnetism in the Huesler Alloys: $Co_2TiSn$ and $Co_2TiAl$. *J. Magn. Magn. Mater.* **1995**, *140–144*, 183–184.

(5) Kanomata, T.; Chieda, Y.; Endo, K.; Okada, H.; Nagasako, M.; Kobayashi, K.; Kainuma, R.; Umetsu, R. Y.; Takahashi, H.; Furutani, Y.; Nishihara, H.; Abe, K.; Miura, Y.; Shirai, M. Magnetic Properties of the Half-Metallic Heusler Alloys $Co_2VAl$ and $Co_2VGa$ under Pressure. *Phys. Rev. B* **2010**, *82*, 144415.

(6) Graf, T.; Winterlik, J.; Müchler, L.; Fecher, G. H.; Felser, C.; Parkin, S. S. P. Chapter One - Magnetic Heusler Compounds. In *Handbook of Magnetic Materials*; Buschow, K. H. J., Ed.; Elsevier, 2013; Vol. 21, pp 1–75.

(7) Pan, Y.; Nikitin, A. M.; Bay, T. V.; Huang, Y. K.; Paulsen, C.; Yan, B. H.; Visser, A. de. Superconductivity and Magnetic Order in the Noncentrosymmetric Half-Heusler Compound ErPdBi. *EPL* **2013**, *104*, 27001.

(8) Müller, R. A.; Desilets-Benoit, A.; Gauthier, N.; Lapointe, L.; Bianchi, A. D.; Maris, T.; Zahn, R.; Beyer, R.; Green, E.; Wosnitza, J.; Yamani, Z.; Kenzelmann, M. Magnetic Structure of the Antiferromagnetic Half-Heusler Compound NdBiPt. *Phys. Rev. B* **2015**, *92*, 184432.

(9) Casper, F.; Graf, T.; Chadov, S.; Balke, B.; Felser, C. Half-Heusler Compounds: Novel Materials for Energy and Spintronic Applications. *Semicond. Sci. Technol.* **2012**, *27*, 063001.

(10) Dorolti, E.; Cario, L.; Corraze, B.; Janod, E.; Vaju, C.; Koo, H.-J.; Kan, E.; Whangbo, M.-H. Half-Metallic Ferromagnetism and Large Negative Magnetoresistance in the New Lacunar Spinel $GaTi_3VS_8$. *J. Am. Chem. Soc.* **2010**, *132*, 5704–5710.

(11) Zhang, X.; Zhang, J.; Zhao, J.; Pan, B.; Kong, M.; Chen, J.; Xie, Y. Half-Metallic Ferromagnetism in Synthetic $Co_9Se_8$ Nanosheets with Atomic Thickness. *J. Am. Chem. Soc.* **2012**, *134*, 11908–11911.

(12) Li, X.; Wu, X.; Yang, J. Half-Metallicity in $MnPSe_3$ Exfoliated Nanosheet with Carrier Doping. *J. Am. Chem. Soc.* **2014**, *136*, 11065–11069.

(13) de Groot, R. A.; Mueller, F. M.; Engen, P. G. van; Buschow, K. H. J. New Class of Materials: Half-Metallic Ferromagnets. *Phys. Rev. Lett.* **1983**, *50*, 2024–2027.

(14) Block, T.; Carey, M. J.; Gurney, B. A.; Jepsen, O. Band-Structure Calculations of the Half-Metallic Ferromagnetism and Structural Stability of Full- and Half-Heusler Phases. *Phys. Rev. B* **2004**, *70*, 205114.

(15) Wu, Y.; Wu, B.; Wei, Z.; Zhou, Z.; Zhao, C.; Xiong, Y.; Tou, S.; Yang, S.; Zhou, B.; Shao, Y. Structural, Half-Metallic and Elastic Properties of the Half-Heusler Compounds NiMnM
12

**Table 1.** Single crystal structure refinement for $Cr_4PtGa_{17}$ at 293 (2) K.

| Refined Formula | $Cr_4PtGa_{17}$ |
| --- | --- |
| F.W. (g/mol) | 1588.33 |
| Space group; Z | $R\overline{3}m$; 3 |
| $a$(Å) | 8.02560 (7) |
| $c$(Å) | 19.6539 (3) |
| V (Å$^3$) | 1096.31 (3) |
| Extinction Coefficient | 0.0029 (1) |
| θ range (º) | 3.109-30.329 |
| No. reflections; $R_{int}$ | 8889; 0.0271 |
| No. independent reflections | 888 |
| No. parameters | 51 |
| $R_1$: $\omega R_2$ ($I>2\delta(I)$) | 0.0098; 0.0243 |
| Goodness of fit | 0.991 |
| Diffraction peak and hole (e$^-$/ Å$^3$) | 0.691; -1.283 |
| Absolute structure parameter | 0.01 (1) |

**Table 2.** Atomic coordinates and equivalent isotropic displacement parameters for $Cr_4PtGa_{17}$ at 293 K. ($U_{eq}$ is defined as one-third of the trace of the orthogonalized $U_{ij}$ tensor (Å$^2$))

| Atom | Wyck. | Occ. | $x$ | $y$ | $z$ | $U_{eq}$ |
| --- | --- | --- | --- | --- | --- | --- |
| Pt1 | 3$a$ | 1 | 0 | 0 | 0.00000 (2) | 0.0052 (1) |
| Ga2 | 9$b$ | 1 | 0.85689 (5) | 0.14311 (5) | 0.92839 (3) | 0.0108 (1) |
| Ga3 | 9$b$ | 1 | 0.47645 (5) | 0.52355 (5) | 0.73822 (4) | 0.0107 (1) |
| Ga4 | 9$b$ | 1 | 0.21087 (5) | 0.78913 (5) | 0.85541 (4) | 0.0111 (1) |
| Ga5 | 9$b$ | 1 | 0.45580 (5) | 0.54420 (5) | 0.97787 (3) | 0.0112 (1) |
| Ga6 | 3$a$ | 1 | 0 | 0 | 0.74997 (6) | 0.0087 (2) |
| Ga7 | 3$a$ | 1 | 0 | 0 | 0.62769 (6) | 0.0089 (2) |
| Ga8 | 9$b$ | 1 | 0.83702 (4) | 0.16298 (4) | 0.79073 (3) | 0.0089 (1) |
| Cr9 | 3$a$ | 1 | 0 | 0 | 0.40569 (8) | 0.0049 (3) |
| Cr10 | 9$b$ | 1 | 0.54107 (6) | 0.45893 (6) | 0.86476 (5) | 0.0049 (2) |

**Table 3.** Anisotropic thermal displacement parameters for $Cr_4PtGa_{17}$.

| Atom | U11 | U22 | U33 | U12* | U13* | U23* |
| --- | --- | --- | --- | --- | --- | --- |
| Pt1 | 0.0052 (1) | 0.0052 (1) | 0.0052 (1) | 0.0026 (1) | 0 | 0 |
| Ga2 | 0.0115 (2) | 0.0115 (2) | 0.0113 (3) | 0.0073 (3) | -0.0020 (1) | 0.0020 (1) |
| Ga3 | 0.0127 (3) | 0.0127 (3) | 0.0100 (3) | 0.0088 (3) | -0.0016 (1) | 0.0016 (1) |
| Ga4 | 0.0079 (2) | 0.0079 (2) | 0.0137 (3) | 0.0011 (2) | 0.0001 (1) | -0.0001 (1) |
| Ga5 | 0.0133 (2) | 0.0133 (2) | 0.0082 (3) | 0.0076 (3) | 0.0019 (1) | -0.0019 (1) |
| Ga6 | 0.0089 (3) | 0.0089 (3) | 0.0083 (5) | 0.0045 (2) | 0 | 0 |
| Ga7 | 0.0100 (3) | 0.0100 (3) | 0.0067 (5) | 0.0050 (2) | 0 | 0 |
| Ga8 | 0.0093 (2) | 0.0093 (2) | 0.0095 (3) | 0.0058 (2) | 0.0005 (1) | -0.0005 (1) |
| Cr9 | 0.0050 (4) | 0.0050 (4) | 0.0048 (7) | 0.0025 (2) | 0 | 0 |
| Cr10 | 0.0049 (3) | 0.0049 (3) | 0.0049 (4) | 0.0023 (3) | 0.0000 (2) | -0.0000 (2) |

*For an explanation of the anisotropic thermal displacement parameters, see *The International Tables for Crystallography*, A. Authier editor, second edition, volume D, pages 231 to 245, John Wiley and Sons, 2014.[82]

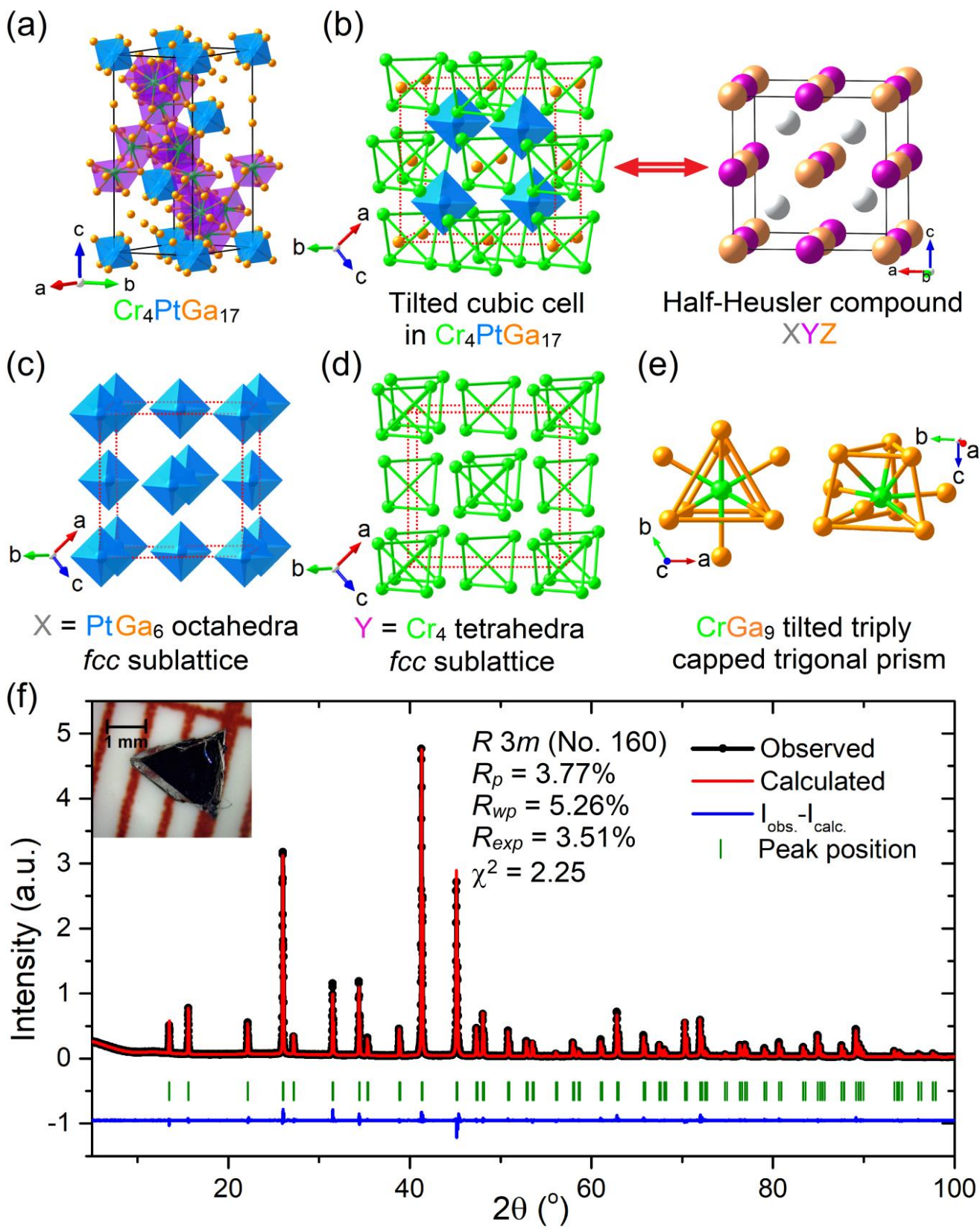

**Figure 1.** (**a**) The crystal structure of Cr$_4$PtGa$_{17}$ where green, blue and orange balls represent Cr, Pt and Ga atoms, respectively. (**b**) (**Left**) Tilted cubic cell in Cr$_4$PtGa$_{17}$; (**Right**) Crystal structure of half-Heusler compound XYZ where grey, purple and orange balls stand for X, Y and Z atoms, respectively. Face-centered cubic sublattices constructed by (**c**) Pt@Ga$_6$ octahedra and (**d**) Cr$_4$ tetrahedra. (**e**) Shape

of $CrGa_9$ polyhedron. Red dotted lines in **(b), (c)** and **(d)** indicate imaginary cubic cell. **(f) (Main panel)** Powder XRD pattern of $Cr_4PtGa_{17}$ with Rietveld fitting. Black line with ball stands for observed pattern. Red line and blue line indicate calculated pattern and difference between observed and calculated patterns, respectively. Vertical green sticks represent Bragg peak positions. **(Inset)** Picture of a $Cr_4PtGa_{17}$ crystal.

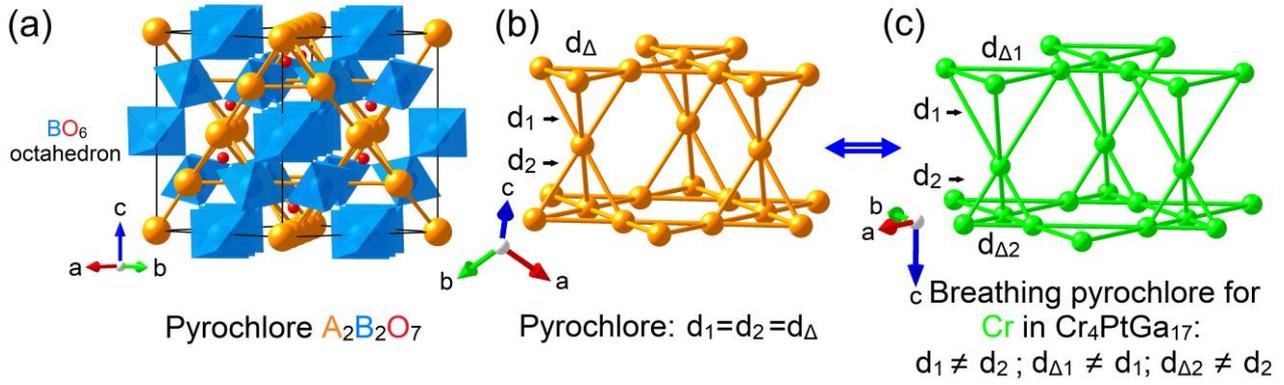

**Figure 2.** (**a**) Crystal structure of an oxide pyrochlore, $A_2B_2O_7$. Orange, blue and red balls represent A, B and oxygen atoms, respectively. The oxygen atoms on the corner of the $BO_6$ octahedra are omitted. (**b**) Normal pyrochlore lattice of the A or B atoms in ideal pyrochlore compounds, where $d_1$ and $d_2$ are the lengths of the sides of two corner-shared tetrahedra and $d_\Delta$ stands for Cr-Cr bond length within the Kagome plane. (**c**) The breathing pyrochlore lattice constructed by Cr atoms found in $Cr_4PtGa_{17}$. Distances $d_1$ and $d_2$ are the separations between Cr atoms for larger and smaller tetrahedra, respectively, while $d_{\Delta 1}$ and $d_{\Delta 2}$ stand for Cr-Cr bond lengths within the tilted Kagome plane for bigger and smaller tetrahedra, respectively.

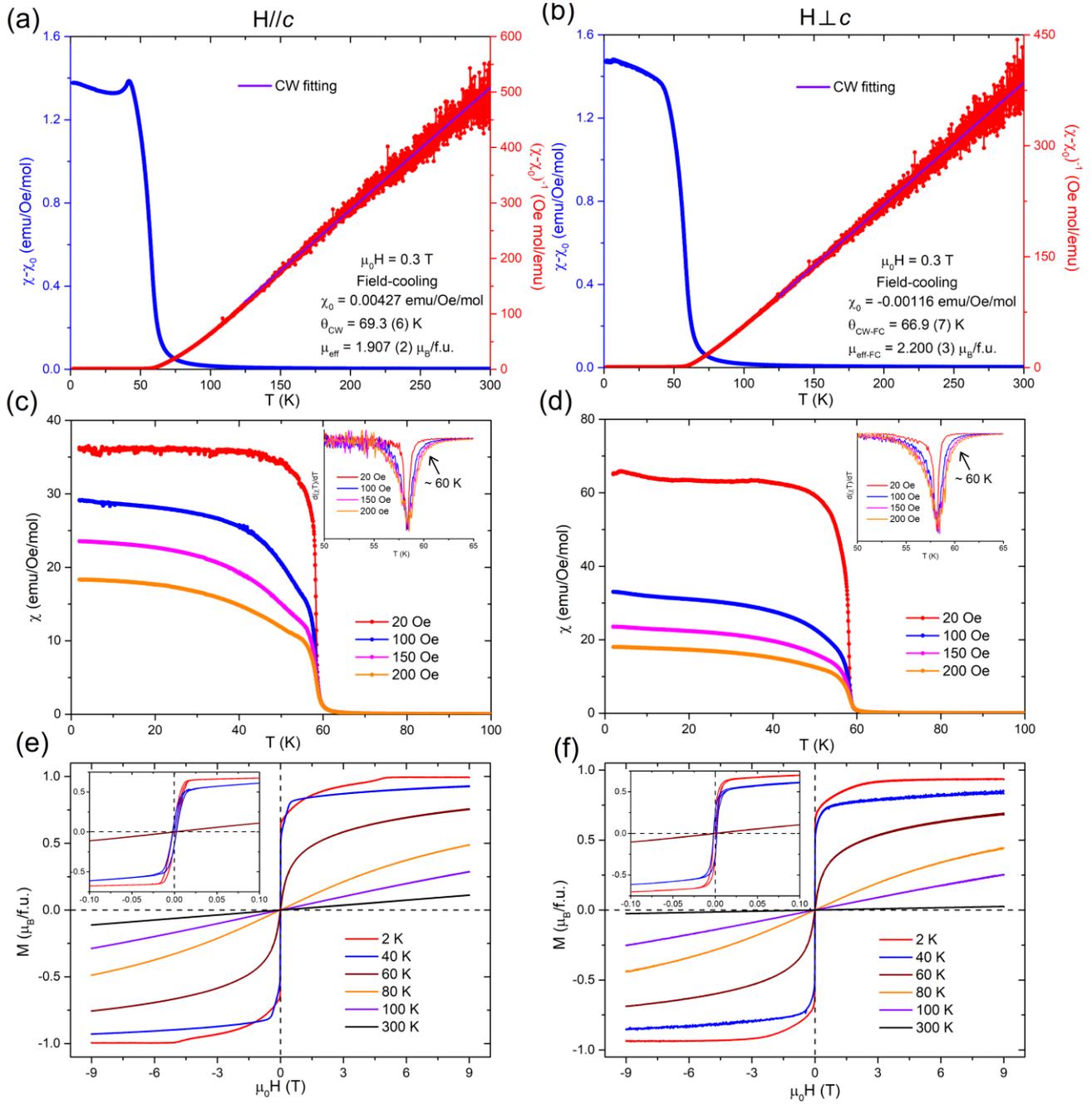

**Figure 3.** Temperature-dependence of magnetic susceptibility ($\chi$) with $\chi_0$ subtracted ($\chi$-$\chi_0$) for Cr$_4$PtGa$_{17}$ crystals from 2 K to 300 K for FC mode when magnetic field of 0.3 T was **(a)** parallel to crystal *c* axis; **(b)** perpendicular to *c* axis. Temperature-dependence of magnetic susceptibility ($\chi$) under zero-field cooling (ZFC) mode from 2 K to 100 K under various magnetic fields which were applied **(c)** parallelly to *c* axis; **(d)** perpendicularly to *c* axis. The **Insets** of **(c)** & **(d)** show the first derivatives of $\chi T$ vs T curves. Hysteresis loops from -9 T to 9 T under various temperatures when magnetic field was applied **(e)** parallelly to *c* axis; **(f)** perpendicularly to *c* axis. The **Insets** of **(e)** & **(f)** illustrate zoom-ins of hysteresis loops from -0.1 T to 0.1 T under 2 K, 40 K and 60 K.

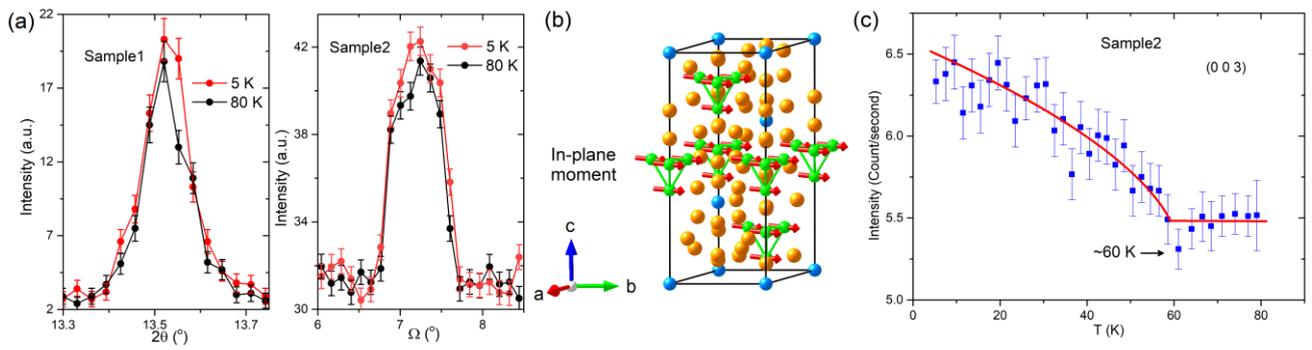

**Figure 4. (a)** Rocking curve scan for Bragg peak (0 0 3) of two different $Cr_4PtGa_{17}$ crystals (samples 1 & 2) at 5 K and 80 K. **(b)** Magnetic structure obtained from single crystal neutron diffraction with the in-plane magnetic moments noted as red vectors. Green, blue and orange balls represent Cr, Pt and Ga atoms, respectively. **(c)** The intensity of the (0 0 3) Bragg peak plotted from 4 K to 80 K for sample 2. Red solid line represents the power-law fit described in the text.

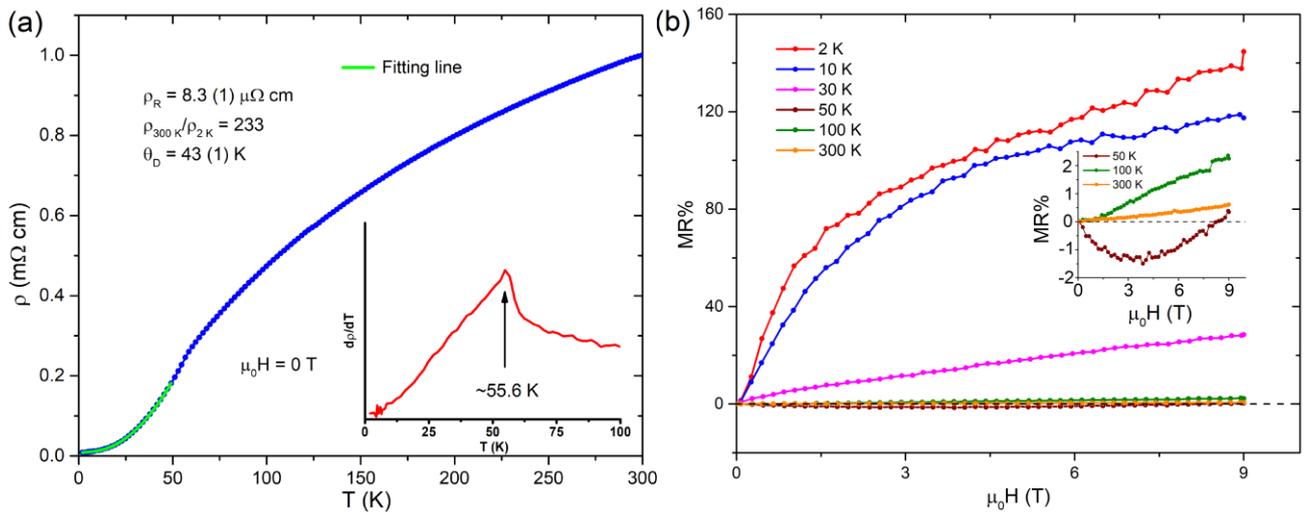

**Figure 5. (a)** (**Main panel**) Temperature-dependence of resistivity for $Cr_4PtGa_{17}$ single crystal under no magnetic field. The green line shows the fitting for a ferromagnetic system. (**Inset**) The first derivative of resistivity *vs* temperature. (**b**) Magnetoresistance from 0 to 9 T at various temperatures.

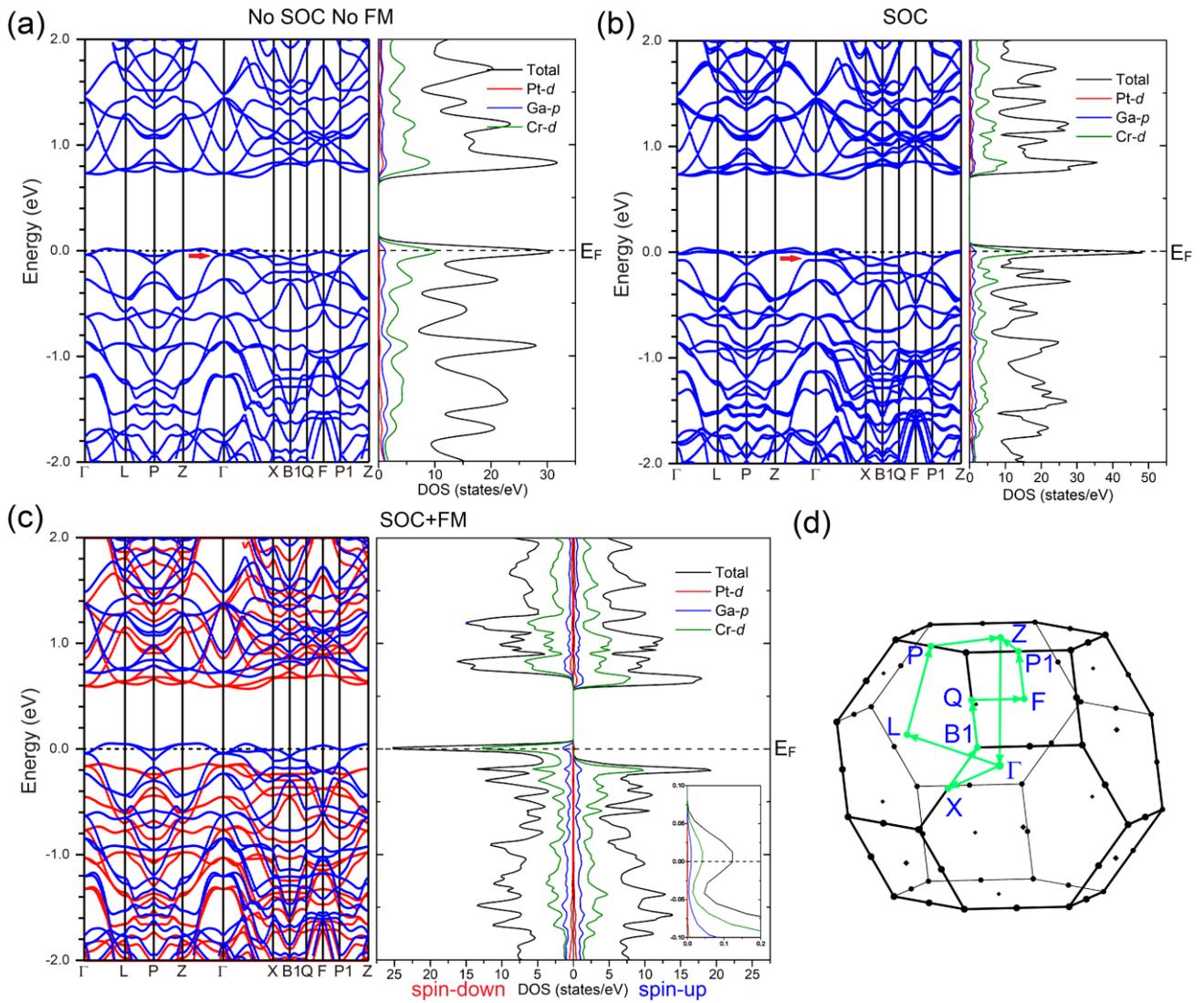

**Figure 6.** Band structure and density of states (DOS) calculations for $Cr_4PtGa_{17}$ **(a)** without including spin-orbit coupling (SOC) and spin polarization (FM); **(b)** considering SOC effects on the Pt atom; **(c)** with inclusion of SOC and FM, where in the band structure, red bands are from spin-down channel and blues bands are from spin-up channel. The inset of DOS figure in **(c)** shows the zoom-in of DOS of spin-up channel from -0.10 eV to 0.10 eV. **(d)** The first Brillouin zone and wave vector path used in all calculations.

**For Table of Contents Only**

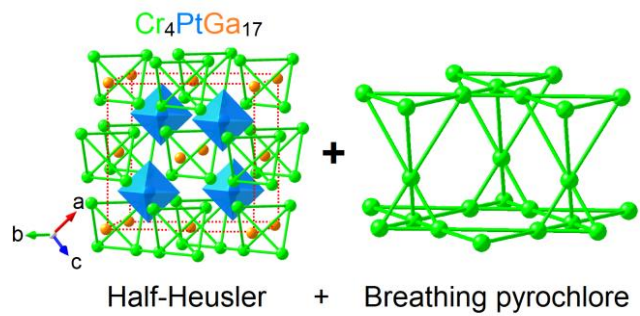

Half-Heusler + Breathing pyrochlore